# Modeling the aging kinetics of zirconia ceramics

Laurent Gremillard[*], Jérôme Chevalier[+], Sylvain Deville, Thierry Epicier, Gilbert Fantozzi,

GEMPPM, INSA de Lyon, 69621 villeurbanne Cedex, France

[*]Material Science Division, LBNL, Berkeley CA 94720, USA

[+]To whom correspondance should be adressed : Jerome.chevalier@insa-lyon.fr

*Abstract*

Yttria-stabilized tetragonal zirconia polycrystals (3Y-TZP) with different microstructures were elaborated. The isothermal tetragonal to monoclinic transformation was investigated at 134 °C in steam by X-ray diffraction, Atomic Force Microscopy (AFM) and optical interferometry. The aging kinetics were analyzed in terms of nucleation and growth, using the Mehl-Avrami-Johnson (MAJ) formalism. Numerical simulation of the aging of zirconia surfaces was also conducted, and the results were used to better fit the aging kinetics. The simulation shows that the exponent of the MAJ laws is controlled not only by the nucleation and growth mechanisms, but also – and mainly – by their respective kinetic parameters. Measurements of nucleation and growth rates at the surface, at the beginning of aging, and the use of numerical simulation allow the accurate prediction of aging kinetics.

*Introduction*

Zirconia ceramics are widely used, particularly for orthopaedic applications such as femoral heads for hip prostheses. Recently the failure of a number of zirconia heads [1,2] draw the attention to the phenomena limiting the life-time of zirconia ceramic pieces, in particular on the aging of zirconia. Aging was first described by Kobayashi et al. in 1981[3]: at room temperature, zirconia is retained in a metastable tetragonal phase by the addition of stabilizing agents (for example yttria); the aging of zirconia consists in a return toward the more stable monoclinic phase. The transformation is martensitic in nature and occurs preferentially at the surface of tetragonal zirconia ceramics. It has been shown that the t-m transformation at the surface of zirconia ceramics is promoted by the presence of water molecules in the environment. Being subject to a volume increase, this t-m transformation induces the formation of microcracks at the surface, and an increase of the roughness. Microcracking leads to a decrease of the mechanical properties[4] ; this could explain the failure of implants after some years in vivo. The aging should then be avoided, or at least kinetics taken into account to calculate the lifetime or zirconia pieces [5,6].

Numerous studies have been conducted to measure the aging kinetics of Y-TZP[4,7,8,14-27] or on Ce-TZP[28] (Yttria or Ceria stabilized Tetragonal Zirconia Polycrystals). Depending on the authors, the laws giving the monoclinic phase amount versus time can be either linear or sigmoidal. Until now, no real effort to rationalize these differences has been made. The most detailed studies [18,21,22,26] have shown that the sigmoidal laws are related to nucleation and growth kinetics (nucleation of the monoclinic phase first on isolated grains on the surface, then propagation to the neighboring grains as a result of stresses and microcracks accumulation). In those cases, the kinetics can be described using Mehl-Avrami-Johnson[29] laws (eq. (1), where the exponent n is of particular interest) :

$$f = 1 - \exp\left(-(b.t)^n\right) \qquad (1)$$

Where f is the monoclinic fraction and t the aging duration. But the exponent n given in or deducted from the literature can vary from 0.3[23] to almost 4[18].

According to Christian [30], these different values of n should suggest different mechanisms. Our paper aims at showing that this exponent n is not only characteristic of the aging mechanism (i.e. nucleation and growth), but also strongly depends on the kinetic features (mainly nucleation speed and interface – or growth – speed).

EXPERIMENTALS

*Materials*

The results concerning 3 batches of zirconia ceramics will be exposed. The starting powder was in each case an ultra-pure Tosoh powder, stabilized in the tetragonal phase with 3 mol % of Yttria and with less than 200 ppm of impurities. The samples were processed by cold isostatic pressing at 300 MPa. For the first batch (noted Z-S$_{1450}$), the compacts were sintered at 1450°C for 5 hours in air (heating and cooling rate: 300°C/h), to achieve grain sizes of approximately 0.5 µm (measured by SEM according to the ASTM E112 standard) and a density of 97% of the theoretical density. For the second batch (noted Z-HIP), Hot Isostatic Pressing (HIP) was conducted after the sintering step in order to achieve full density ( > 99 %) and the occurrence of less pores and defects. For the third batch (noted Z-S$_{1600}$), the compacts were sintered at 1600°C for 5 hours in air to achieve 1 µm grain size.

*Aging*

The Low Temperature Degradation was quantified by following the monoclinic content on the surface of the material versus aging time. The monoclinic content, $X_m$, was measured by X-ray diffraction (XRD) and calculated from the Garvie and Nicholson [31] equation (2):

$$X_m = \frac{I_m\{\bar{1}11\} + I_m\{111\}}{I_m\{\bar{1}11\} + I_m\{111\} + I_t\{101\}} \quad (2)$$

Where $I_J\{hkl\}$ is the area of the {hkl} peak of the phase j measured by XRD. The X-ray penetration depth (and then the depth of the analyzed area) is around 5 μm [32].

The samples were polished with diamond pastes down to 1 µm so as to achieve a perfectly smooth surface (roughness less than 4 nm measured by the mean of an optical interferometer, Phase Shift Technology®) and to remove all surface stresses due to machining. The monoclinic phase was undetectable before aging tests.

Accelerated aging tests were conducted in steam, at 134 °C, under a pressure of 2 bars (this corresponds to a standard sterilization procedure). It has been shown that one hour of this treatment corresponds roughly to four years at body temperature [33].

The optical interferometer was also used to take images of the surface of aged samples, in order to follow the surface changes during aging, in particular to investigate the appearance of monoclinic nuclei and their growth. A quantification of the number and heights of monoclinic nuclei was conducted on these images.

AFM experiments were performed on a D3100 microscope (Digital Instruments™) in contact mode, using oxide sharpened silicon nitrided probes, with an average scanning speed of 20 µm.s-1. AFM allowed the observation of the very first stages of monoclinic spots growth, thanks to its improved lateral resolution as compared to optical interferometry, confirming thus the interferometry measurments.

EXPERIMENTAL RESULTS

The relationship between the amount of monoclinic phase and aging time for the three zirconia ceramics is shown in Fig. 1. The curves can be considered as nearly sigmoïdal for Z-$S_{1450}$ and Z-HIP and almost linear with a plateau for Z-$S_{1450}$. The n exponent which can be derived from the slope of the $\ln(\ln(1/(1-f)))$ versus $\ln(t)$ plot of Fig.2 varies from 1 to 4 for the three batches (see Fig.3). According to the MAJ theory, this could suggest different mechanisms. Also the n exponent is not constant over the overall kinetic. Recent works suggest that aging occurs in zirconia by a nucleation and growth mechanism[18,21]. Clear evidence for this mechanism is shown by the AFM pictures of Fig.4.

Discuter sur le fait que dans cette nuance on a peu de nucléation par rapport à 3Y-TZP ??? (cf la discussion qu'on a eu, Jérôme)

The figure shows the appearance of monoclinic nuclei, then their growth versus time in one of the tested zirconia ceramic. This was observed for the three zirconia ceramics, but with different kinetics. In agreement with a previous study [18], we observed that the nucleation rate (number of new monoclinic regions appearing by time and by unit surface area) and the growth rate (increase of the diameter and the height of a given monoclinic region by time) were nearly constant, at least during the first hours aging. Their values are compiled in table 1. It is seen from these results that the nucleation rate is strongly influenced by the grain size (the higher the grain size, the higher the nucleation of new monoclinic regions) and the density (even small defects acting probably as nucleation sites). These results show that aging kinetics are strongly affected by the microstructure and confirm that decreasing the grain size and increasing the density by Hot Isostatic Pressing is a good way to decrease the aging kinetics of zirconia.

MODELS

*Mehl-Avrami-Johnson (MAJ) laws for nucleation and growth.*

The analytical analysis conducted on MAJ law applied to aging of zirconia can be found in more details in ref. 18. Figure 5 shows schematically the geometry of the transformed spots appearing on the surface in zirconia. From the current experimental results and previous works [18-21], it is recognized that aging occurs by a constant nucleation of new sites with time and an increase in diameter and height of existing ones at constant rate (providing their is enough space to growth). This can be summarized in a mathematical form [18]:

$$d_\tau(t) = \alpha_d \cdot (t - \tau) \qquad (2)$$

$$h_\tau(t) = \alpha_h \cdot (t - \tau) \qquad (3)$$

$$dN = N_r \cdot (1 - f) \cdot d\tau \qquad (4)$$

where $d_\tau(t)$ and $h_\tau(t)$ represent the diameter and the height of a spot originating at time $t = \tau$, and where dN corresponds to the number of nuclei created per unit area between $\tau$ and $\tau + d\tau$. f is the monoclinic fraction. $\alpha_d$ and $\alpha_h$ are constants characteristic of the diameter and height growth rates and $N_r$ the nucleation rate per unit time and area.

The surface uplift observed at the surface corresponds to the volume expansion due to t-m transformation, known to be accompanied by a volume expansion of about 4% [34], thus by a linear expansion, k, of about 1.3%. Thus, the volume of a given monoclinic region originating at $t = \tau$ is given by :

$$v_\tau(t) = \frac{\pi}{12} \cdot \frac{\alpha_d^2 \cdot \alpha_h}{k} \cdot (t - \tau)^3 \qquad (5)$$

The volume of monoclinic phase created at time t from nuclei originating between $\tau$ and $\tau + d\tau$ is : $dVm = v_\tau(t)\,dN$. If we consider a thickness of material $l = 5$ µm to compare with X-ray diffraction results[32], the increase in monoclinic fraction is :

$$df = \frac{dVm}{l} = v_\tau(t) \cdot dN \qquad (6)$$

Integrating eq. (4) and (5) in (6) gives :

$$\frac{df}{(1-f)} = \frac{\pi \cdot \alpha_d^2 \cdot \alpha_h \cdot N_r}{12 \cdot k \cdot l} \cdot (t - \tau)^3 \cdot d\tau \qquad (7)$$

The monoclinic fraction is then obtained by integration of equation (7):

$$f = 1 - \exp\left[-\left(\frac{p \cdot a_d^2 \cdot a_h \cdot N_r}{48 \cdot k \cdot l}\right) \cdot t^4\right] \quad (8)$$

This MAJ type kinetic was successfully applied to one zirconia in ref 18 and can apply well to fit the results of Z-HIP and moderately of Z-S$_{1450}$ (see Fig. 1). However, this equation considers that the n exponent is constant, equal to 4, while experimental results show a variation of n with time. Moreover, this equation does not apply to Z-S$_{1600}$, where the value of n is around 1. In the case of this latter material, the ratio of nucleation rate on growth rate is much more important (see table 1), so that most of the monoclinic fraction created during aging is due to nucleation. The interaction between nuclei occurs rapidly, which hinders their growth. This is not taken into account in this analysis (in equations 2 and 3).

Two other factors can affect the rightness of this analysis. The most important is probably the fact that the nuclei have already a size when they appear[35,36] (they are not infinitely small as considered in the previous calculus). Also the X-Ray absorption profile is not considered. All these problems can be solved with a numerical simulation.

*Numerical simulation*

In order to model the X-Ray diffraction results obtained in the three materials, a numerical simulation was conducted[37]. The program we set up provides a calculation of the transformed fractions versus time, knowing the nucleation rate, the growth speed and the initial size of the nuclei (Nr, $\alpha_h$ or $\alpha_d$, and H or D). It uses only a few simple rules:

- The nuclei are cone-shaped;
- the number of new nuclei during a given time interval dt is proportional to dt, to the non-transformed surface and to the initial nucleation speed;

- the new nuclei are distributed randomly on the non-transformed surface (if the nuclei arrives on an already transformed area, the software will put it on another location);

- the growth speed $\alpha_d$ is constant, and during each dt, all the point inside a given radius ($\alpha_d$ x dt/2) from the previously transformed areas are transformed;

- the transformed depth under a point of the surface is proportional to the growth speed and to the time from the first transformation;

- the X-ray absorption in the material is considered, so that the curves provided by the program can directly be compared to the aging kinetics measured by X-ray diffraction.

Each time step in the simulation is divided in a nucleation calculus, then a growth calculus. The initial state of the simulated material is completely 'transformation free', and its whole surface is susceptible to transformation.

An example of images of the surface calculated by means of this simulation is given on Fig. 6.

EXPERIMENTAL VALIDATION OF THE NUMERICAL MODEL

Fig. 1 provides a comparison between the experimental X-Ray diffraction results and the numerical simulation conducted from the parameters of table 1. In the simulations, D was taken as the grain size (i.e. 0.5 µm for Z-S$_{1450}$ and Z-HIP or 1 µm for Z-S$_{1600}$). They show in each case a good agreement. From the simulated laws, we can derive an n exponent which could be used to fit these curves by a MAJ law: by tracing $\dfrac{\Delta\left(\ln\left(\ln\dfrac{1}{1-f(t)}\right)\right)}{\Delta(\ln(t))}$ versus time (fig. 7), we obtain a curve with a plateau giving the value of the exponent n. This figure leads us to a few commentaries. First, the value of n is ranging between 1 and 4. Second, this value is not constant during aging; this is coherent with what is observed experimentally: indeed the figure 3 shows the same evolution of $\dfrac{\Delta\left(\ln\left(\ln\dfrac{1}{1-f(t)}\right)\right)}{\Delta(\ln(t))}$ versus time (ascending phase, plateau and decrease) for the experimental measures.

DISCUSSION

In order to understand the influence of the kinetic parameters ($\alpha_h$, $\alpha_d$ and $N_r$) on the shape of the sigmoidal laws (in other words on the value of the exponent n), a set of more than 200 simulations was conducted. From all the results, the exponent n is calculated according to the method previously exposed. The figure 8 shows that this exponent is strongly related to the proportion of new transformed phase appearing by nucleation compared to this appearing by growth of existing nuclei. The former quantity is proportional to the nucleation rate $N_r$ and to the square of the monoclinic spots initial diameter (D²). The later quantity is proportional to the growth speed ($\alpha_h$). Thus the pre-cited proportion is proportional to $\dfrac{\alpha_d}{N_r D^2}$. Assuming that the monoclinic spots initial diameter is the grain size, the experimental data seem to follow the tendency given by the simulation. This figure 8 may also allow us to explain the evolution of n during an aging kinetic (fig. 2). We may explain the lower value of n at the beginning of aging by a stronger influence of the nucleation (equivalent to a small value of $\dfrac{\alpha_d}{N_r D^2}$). Indeed, the first nuclei are small, and their growth produces less transformed material than the growth of older nuclei. On the plateau, the proportion of transformed material appearing by nucleation and by growth is constant. The decrease and peak of the simulated n near the end of the aging curves is due to side effects (as the software does not work with an infinite material, but with an area of 2048x2048 points). This analysis is confirmed by the fig. 9. This figure plots the evolution of n during some simulations on the material Z-P versus the proportion of new monoclinic phase created by growth and by nucleation (dfg/dfn). This curve shows that the variation of n vs dfg/dfn inside a simulation (ie for a same material) is of the same kind as the one shown on fig. 8.

From all these results, it is obvious that only numerical simulation allows to fit the aging kinetics, whatever is the ratio of nucleation on growth rate. Numerical simulation can be a powerful in terms of lifetime prediction, since recording the initiation of aging (nucleation and growth rate at the very beginning of transformation) by the means of AFM or optical interferometer allow to predict the overall aging kinetics with excellent accuracy.

*Conclusion*

This study aimed at clarifying the nucleation and growth kinetics of zirconia ceramics aging. It has been shown that in some cases Mehl-Avrami-Johnson laws with an exponent 4 were a good analysis of this phenomenon. But in most cases, the exponents calculated from the data were widely inferior to 4, typically between 0.3 and 3.5. Numerical simulations allowed us to explain this fact. They showed that:

- the MAJ laws exponent is closely related to the ratio between nucleation rate and growth speed;
- this exponent is not constant during the aging (which is consistent with experimental data), and its evolution can be related once more to the proportion of transformed material appearing by nucleation and by growth.

Since a same mechanism can achieve very different values for n, we can then conclude that the knowledge of the MAJ laws exponent is in itself far from enough to characterize the nucleation and growth mechanism. The knowledge of nucleation rate and growth rate at the very beginning of the transformation can allow to simulate correctly the transformation kinetics. This is particularly crucial for predictions of aging at body temperature.

In terms of impact of the microstructure on aging, it is confirmed here that decreasing the grain size and the density plays a beneficial role on nucleation rate, and thus on aging resistance.

We believe that the reasoning we used here must not be limited to the study of zirconia ceramics aging. This kind of simulation can help to study every phenomenon comprising nucleation-and-growth kinetics.

| Batch | Z-S$_{1450}$ | Z-HIP | Z-S$_{1600}$ |
|---|---|---|---|
| **Nucleation rate (Nr) (nuclei/h/µm$^2$)** | $8.10^{-3}$ | $3,7.10^{-4}$ | $2.10^{-2}$ |
| **Diameter Growth rate α$_d$ (µm/h)** | 2,4 | 2,7 | 0,8 |
| **Height Growth rate α$_h$ (nm/h)** | 24 | 27 | 8 |
| **α$_d$/N$_r$** | $4,8.10^2$ | $7,3.10^3$ | 40 |
| **n** | 3,5 | 4 | 1,4 |

*Table 1 : Nucleation and growth parameters (Nr, α$_d$, α$_h$) and n values for the three zirconia ceramics*

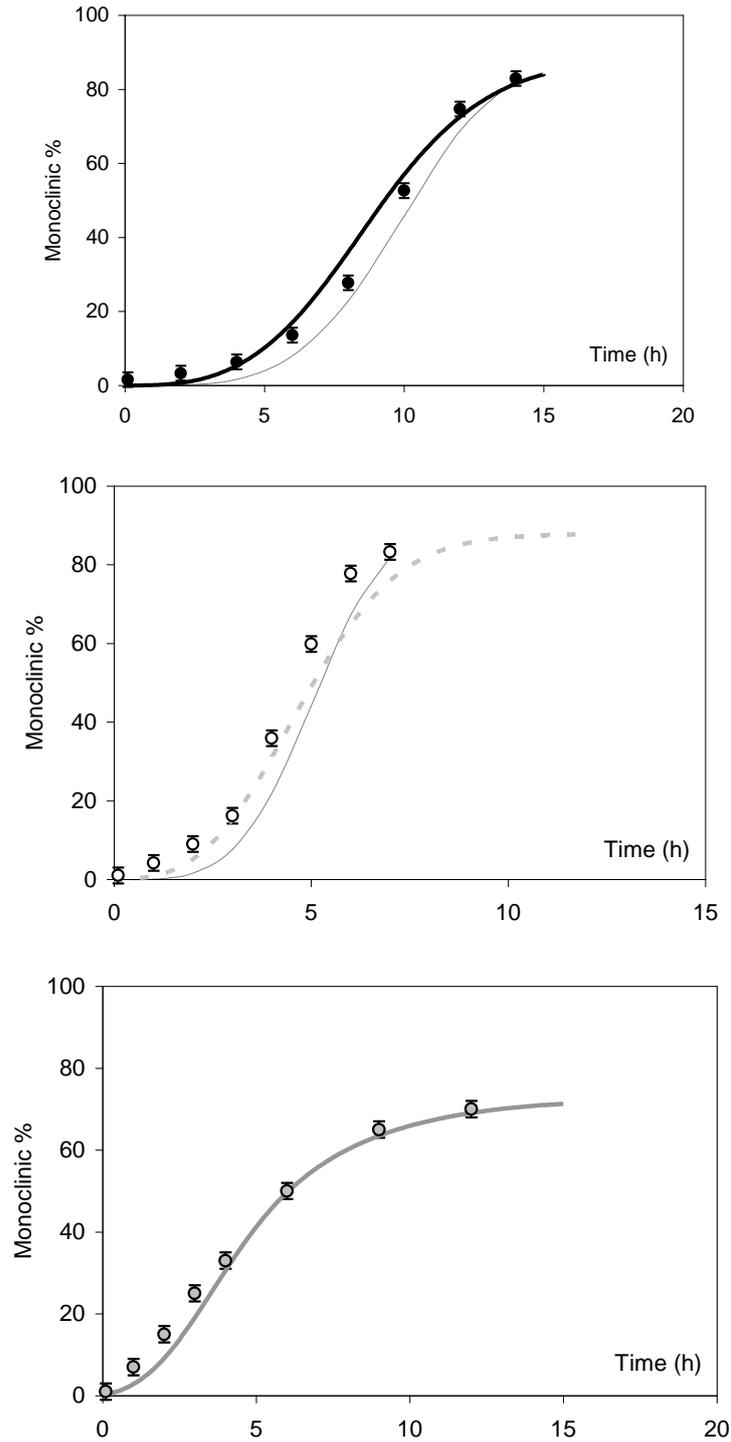

Figure 1: Comparison between experimental points, MAJ analytical model calculated by Chevalier et al. (thin dotted line) and simulated aging kinetics (heavy lines) for the Z-HIP (black), Z-S$_{1450}$ (white) and Z-S$_{1600}$ (grey) materials.

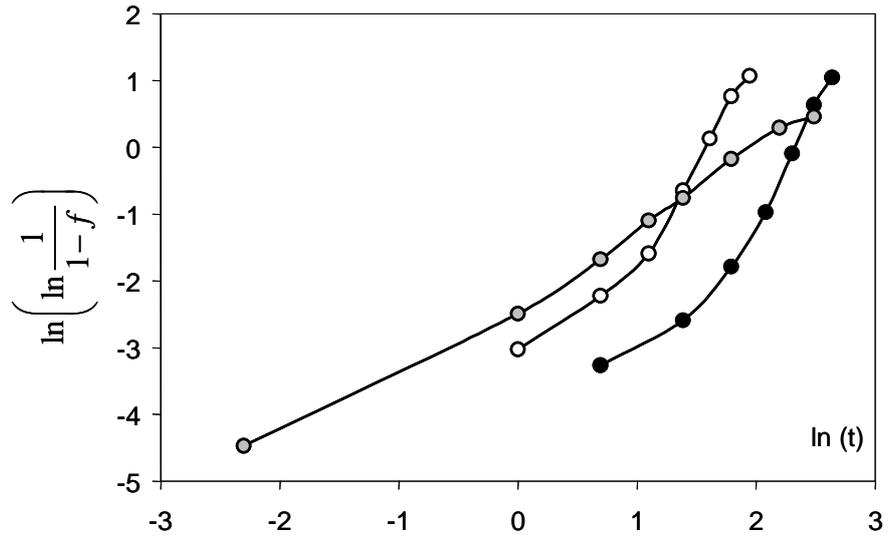

Figure 2: Mehl-Avrami-Johnson plot of the experimental data (black: Z-HIP; white: Z-S$_{1450}$; grey: Z-S$_{1600}$). The slope of each curve should give the value of the exponent n in the MAJ laws.

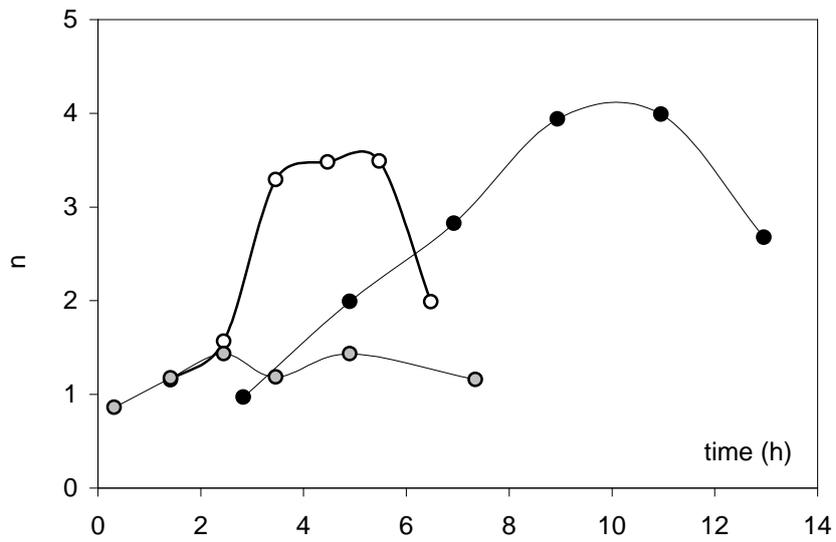

Figure 3: Variation of the exponent n with time during the aging kinetics (black: Z-HIP; white: Z-S$_{1450}$; grey: Z-S$_{1600}$).

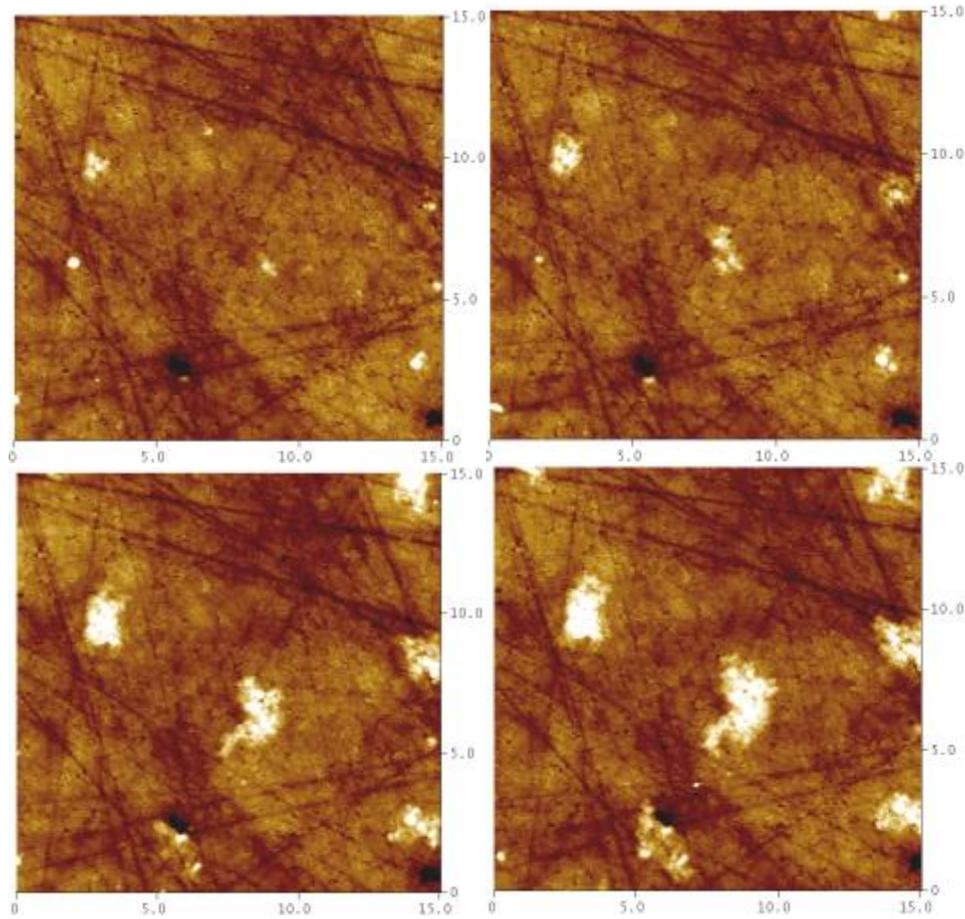

Figure 4: AFM height images of transformation at the surface of zirconia ceramics (vertical scale contrast: 50 nm).

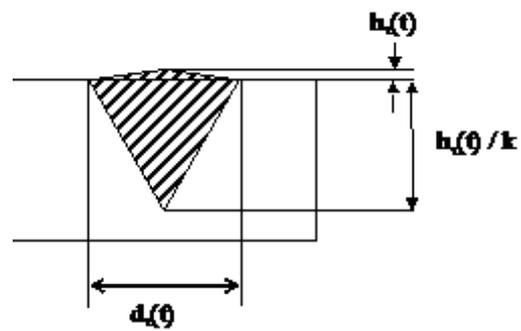

Figure 5: geometry of a transformed nucleus, side view (left) and extension of the transformation, top view (right)

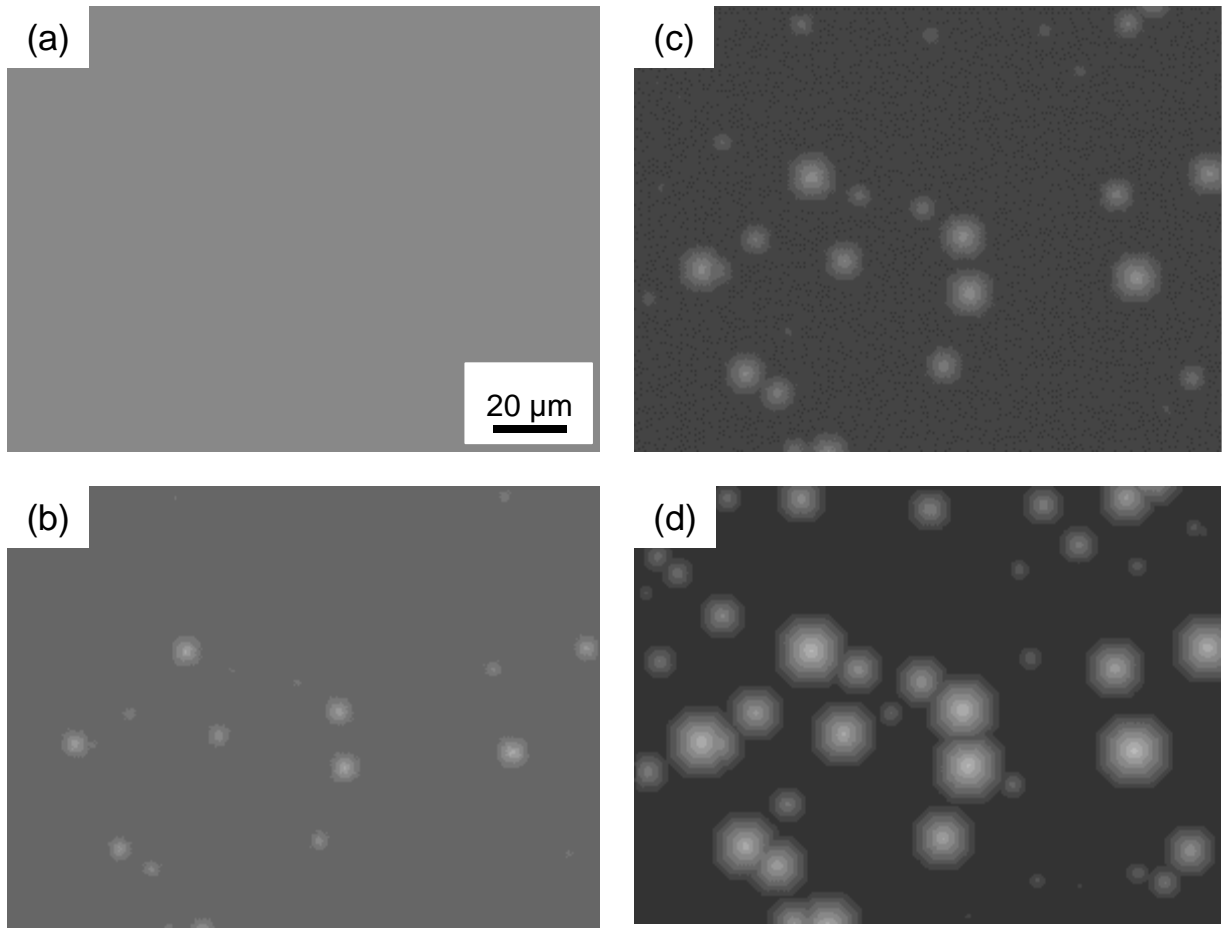

Figure 6: Simulated images of the surface of Z-HIP after 0h (a), 3h (b), 5h (c) and 7h(d).

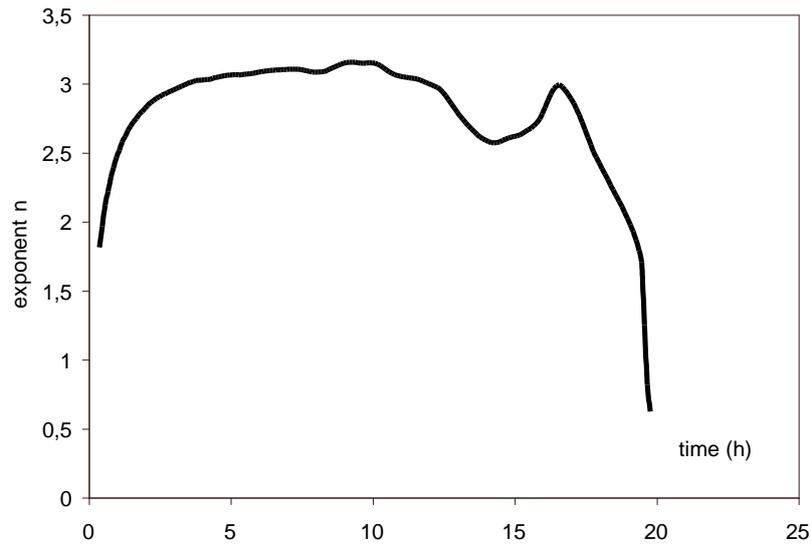

Figure 7: Variation of the exponent n with time during a simulated aging kinetic for Z-HIP.

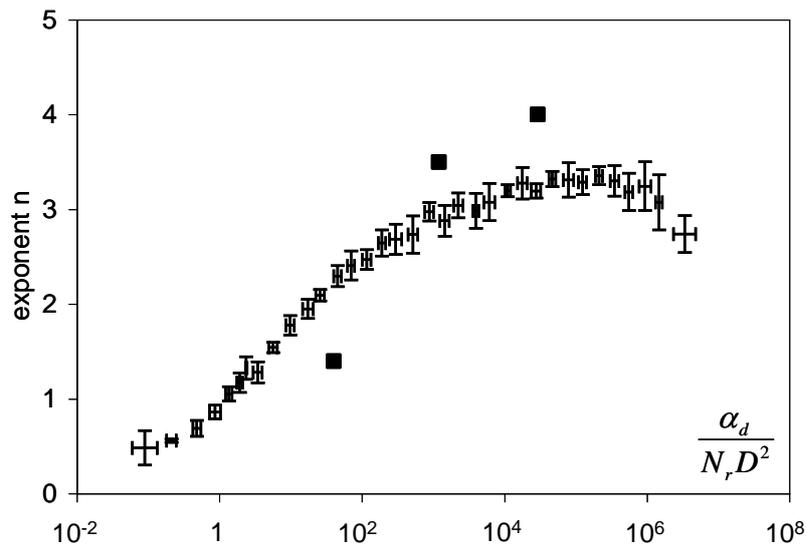

Figure 8: Variation of the exponent n with the kinetic parameters; the black squares indicate the experimental points.

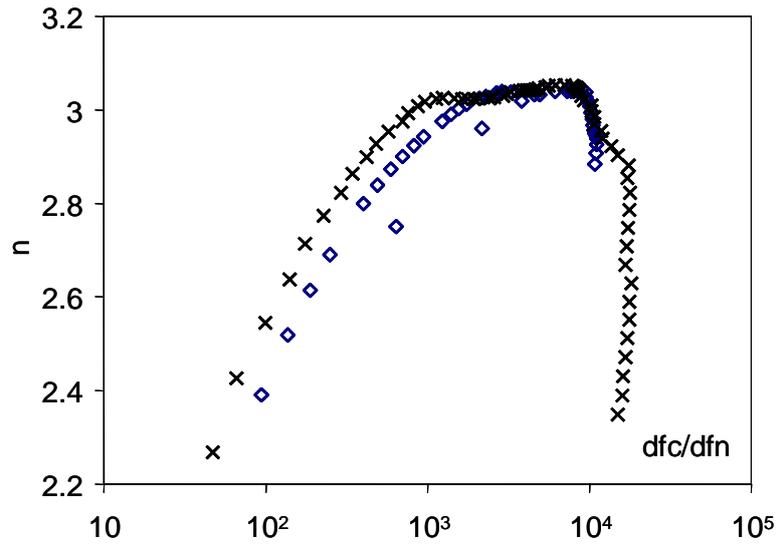

Figure 9: Variation of the exponent n with the proportion of new phase appearing by growth (dfc) and by nucleation (dfn). This is derived from two simulations for Z-HIP.